\documentstyle[12pt]{article}

\def\beq{\begin{equation}}   \def\eeq{\end{equation}}

\newcommand{\gsim}{\lower.7ex\hbox{$\;\stackrel{\textstyle>}{\sim}\;$}}
\newcommand{\lsim}{\lower.7ex\hbox{$\;\stackrel{\textstyle<}{\sim}\;$}}

\newcommand{\ra}{\rightarrow}

\begin{document}

\def\lsim{\mathrel{\rlap{\lower3pt\hbox{\hskip0pt$\sim$}}
    \raise1pt\hbox{$<$}}}         %less than or approx. symbol
\def\gsim{\mathrel{\rlap{\lower4pt\hbox{\hskip1pt$\sim$}}
    \raise1pt\hbox{$>$}}}         %greater than or approx. symbol

\begin{titlepage}
\renewcommand{\thefootnote}{\fnsymbol{footnote}}

\begin{flushright}
CERN-TH/96-356\\
TPI-MINN-96/26-T\\
UMN-TH-1521/96\\
hep-th/9612128\\

\end{flushright}

\vspace{0.3cm}

\begin{center}
\baselineskip25pt

{\Large\bf 
 Domain Walls in Strongly Coupled Theories}

\end{center}

\vspace{0.3cm}

\begin{center}
\baselineskip12pt

\def\thefootnote{\fnsymbol{footnote}}

{\large G. Dvali}

\vspace{0.1cm}
Theory Division, CERN, CH-1211 Geneva 23, Switzerland
\vspace{0.2cm}

{\em and}

\vspace{0.3cm}
{\large  M.~Shifman} 

\vspace{0.1cm}
Theory Division, CERN, CH-1211 Geneva 23, Switzerland \\

and \\

  Theoretical Physics Institute, University of Minnesota, Minneapolis,
MN 54555 USA$^\dagger$ \\[0.5cm]

\vspace{0.7cm}

{\large\bf Abstract} \vspace*{.25cm}
\end{center}

Domain walls in strongly coupled  gauge theories are discussed.
A general mechanism is suggested  automatically leading to   massless gauge 
bosons
localized on  the wall. In one of the models considered,
outside the wall the theory is in the non-Abelian confining phase,
while inside the wall it is in the Abelian Coulomb phase.
 Confining property of
the non-Abelian theories is a key ingredient of the
 mechanism which may be of practical use in the context of
the dynamic compactification scenarios. 

In supersymmetric ($N=1$)
Yang-Mills theories the energy density of the wall can be
{\em exactly} calculated  in the strong coupling regime. This calculation 
presents a further example of non-trivial physical quantities
that can be found exactly by exploiting specific properties of supersymmetry.
A key observation is the fact that the wall in this theory is a BPS-saturated 
state.

\vspace{1.3cm}

\hfill

\begin{flushleft}
CERN--TH/96--356\\

December 1996

\vspace{0.25cm}

\rule{2.4in}{.25mm} \\
$^\dagger$ Permanent address.

\end{flushleft}

\end{titlepage}

\newpage

{\em 1. Introduction} \hspace{1cm}  Domain
walls are inherent to field theories with spontaneously broken 
discrete symmetries. This phenomenon -- occurrence of the domain walls --
is quite common in solid state physics. In high energy physics, in many 
models,  such  symmetries are present too, for instance, a discrete
symmetry  associated
with $CP$. Domain walls then naturally appear in course of evolution 
of 
our Universe and must be taken into account  in cosmological considerations. 
Discussion
of the issue started over twenty years ago \cite{ZKO}. Recently 
domain walls in supersymmetric (SUSY) theories
(along with other topological or non-topological defects) were proposed
as a possible mechanism for dynamical compactification
which, simultaneously, ensure spontaneous SUSY breaking \cite{DS}.
Within this approach, the matter our world is built from
is nothing but the  zero modes
localized on the wall. 

In this Letter we consider new classes of the domain walls
which appear in the strongly coupled gauge theories. The first model is an 
example of a domain wall which traps massless gauge particles on the wall.
Localizing the gauge bosons in the core of the defects  is important for 
building realistic phenomenology based on the 
dynamical compactification scenarios \cite{DS}. Spinless bosons and spin-1/2
fermions, whose interactions are properly arranged, admit  localized zero 
modes on the wall, a well-known fact. To the best of 
our knowledge, no fully satisfactory mechanism was suggested for localizing 
the massless vector
gauge fields so far although 
 attempts in this direction were reported
in the literature (e.g.  \cite{oleg}).

The second example of an unusual domain wall is specific to SUSY gauge
theories. This particular wall is not suitable for dynamical compactification.
Nevertheless, we find this object extremely interesting since 
it has a remarkable property.  Although the wall appears in the
strong-coupling theory, its energy density per unit area $\varepsilon$ is
{\em exactly} calculable. In a sense, the situation reminds
that with the monopole mass in the Seiberg-Witten solution \cite{SW},
although we deal with $N=1$ supersymmetry, not $N=2$. Explicit calculation 
of $\varepsilon$, to be carried out below,  helps clarify one old 
question in supersymmetric Yang-Mills theory, which is in the focus of the 
ongoing polemics. It is known that the supersymmetric gluodynamics
possesses $Z_{2T(G)}$ symmetry, a remnant
of the anomalous $U(1)$ \cite{Witten1}. (Here $T(G)$ is the Dynkin index
for the adjoint representation of the gauge group $G$, normalized in such
a way that for $SU(N)$ the index $T(SU(N)) = N$.) A controversy
continues as to whether this $Z_{2T(G)}$ is spontaneously broken in a
standard way, implying the conventional domain walls, or there is
an additional superselection rule, implying that the vacuum angle
$\theta$ varies not from 0 to $2\pi$, as  usually believed, but, rather
from 0 to $2\pi T(G)$. Arguments {\em pro} and {\em contra} were given;
they are  summarized  in Ref. \cite{Smilga1}.  Our result, confirming
a finite 
value of $\varepsilon$, favors the first option.

\vspace{0.2cm}

{\em 2. Massless gauge fields on the wall} \hspace{1cm} A mechanism we 
suggest does not depend on whether or not the theory at hand is 
supersymmetric. To elucidate the idea we will consider a simple 
(non-supersymmetric) example. Assume we have $SU(2)$ Yang-Mills theory 
whose 
matter
sector contains:

(i) one left- and one right-handed
fermion doublet field $(\psi_{L})_{\alpha}$ and $(\psi_{R})_{\alpha}$ in the
fundamental representation of 
$SU(2)$,
one scalar field $\chi^a$ in the adjoint representation,
and one real scalar field $\eta$ carrying no color indices. The interaction 
Lagrangian has the form
$$
{\cal L} = -\frac{1}{4g^2}G_{\mu\nu}^aG_{\mu\nu}^a
+ \bar{\psi_L}\not\! \! D\psi_L + \bar{\psi_R}\not\! \! D\psi_R - \left (
h\eta\bar{\psi_L}\psi_R + {\rm h.c.}\right )
+ $$
\beq
\frac{1}{2}(D_\mu \chi^a)^2 
- \frac{1}{2}\lambda' (\chi^2 + \kappa^2  - v^2 + \eta^2 )^2 
+\frac{1}{2}(\partial_\mu\eta )^2
-\lambda(\eta^2 - v^2)^2\, ,
\label{Lagrstrc}
\eeq
where $G_{\mu\nu}^a $ is the gluon field strength tensor, $v$  and $\kappa$
are positive 
parameters of dimension of mass assumed  to be
much larger than the scale parameter $\Lambda$ of the $SU(2)$ gauge theory 
at hand
\footnote{More exactly, we will assume that
$\lambda v^2\ll \Lambda^2$ but $\sqrt{\lambda} v^2\gg \Lambda^2$.
The latter requirement is introduced for simplicity.
If this requirement is imposed we can ignore
the shift of the vacuum energy due to the gluon condensate outside
the wall.}, $\lambda$ and  $\lambda '$ are (small) dimensionless coupling 
constants, $g$ is the 
gauge coupling constant.

The theory is obviously $Z_2$ invariant under the transformation
$\eta \ra -\eta$ and $\psi_L \ra i\psi_L,~~~ \psi_R \ra -i\psi_R$.
In the true vacuum of the theory this symmetry is spontaneously 
broken, and the field $\eta$ develops a vacuum expectation value (VEV),
\beq
\eta = \, v \,\,\, \mbox{or} \,\,\, - \, v \, .
\label{2v}
\eeq
Correspondingly, the self-interaction potential for $\chi$ is stable, and the
gauge $SU(2)$ is not spontaneously broken. The theory is in the confining 
phase. All observable degrees of freedom are bound states of
gluons and/or matter fields, with masses 
\footnote{In the $SU(2)$ gauge theory with one quark flavor there is
an ``accidental" global $SU(2)$ symmetry due to the fact that
anti-doublet in $SU(2)$  is the same as doublet. If this global symmetry is
spontaneously broken we might get massless ``pions". This is 
not a serious problem, however. To avoid massless ``pions" suffice it
to consider gauge $SU(3)$. Another possibility is to assume that $hv\sim 
\Lambda$, in which case the would-be pions acquire a mass of order
$\Lambda$.}
of order of $\Lambda$. The mass of 
the
$\eta$ quantum is of order 
\beq
m =\sqrt{2\lambda}\,  v \, .
\label{em}
\eeq

The theory has a stable domain wall interpolating between two different 
vacua
in Eq. (\ref{2v}),
\beq
\eta_0 = v \,{\rm tanh}\, (mz) \, .
\label{wall}
\eeq
For definiteness  the wall is placed in the
$\{x,y\}$ plane; the width of the wall in the $z$ direction is
of order of $m^{-1}$.

Let us consider gauge-non-singlet
massless modes localized on the wall. First,
there are two massless fermion doublet modes
localized on the membrane,
\beq
(\psi_{L,R})_{\alpha} = (\nu _{L,R})_\alpha\,
{\rm  e}^{-2h\int_0^z\eta_0(z')dz'}
\eeq
where 
$\nu$ depends only on $x,y$ and $t$, and 
$\gamma_z\nu = \nu$. Localization of these modes is due to the
index theorem \cite{index} and has nothing to do  with the gauge dynamics.
This is simply because of the topologically non-trivial boundary
conditions on the fermion mass,   $m_{\psi}(- \infty) = - m_{\psi}(+\infty) =
- hv$. The localization scale  is governed by $h$ and becomes infinite
if $h \ra 0$. One of the crucial observations of the present work
is that the  gauge-charged fermions (or scalars),
as well as massless gauge fields  can stay localized on the wall
even in the limit $h = 0$ due to  confining gauge dynamics outside
the wall (see below).

To see that this is indeed the case consider the behavior of the $\chi$ in the
classical wall background. As was mentioned, far away from the wall, when 
$\eta$ is close to $v$,
the self-interaction potential for $\chi$ is  stable  and
there is no spontaneous breaking of the gauge $SU(2)$.
However, inside the wall $\eta\approx 0$, and the
self-interaction 
potential for $\chi$ becomes unstable. It is not difficult to check that
for the wide rang of parameters $\chi$ becomes tachionic in the core
and develops a vacuum
expectation value $\chi^2 \sim v^2$. 

Indeed, consider 
\footnote{The line of reasoning  we follow 
here 
is analogous to that
of Ref. \cite{string} for  superconducting cosmic strings.}
a linearized equation for small perturbations in
$\chi^a = \delta_{3a}\chi_0{\rm e}^{-i\omega t}$
in the  kink background (\ref{wall}) \beq
\left\{ -\partial_z^2 + \lambda '\left[ \kappa^2 + v^2({\rm tanh}^2(mz) - 
1)\right]
\right \} \chi_0 = \omega^2 \chi_0\, , 
\label{arastab}
\eeq
This equation, say, for $\kappa = 0$ is a one-dimensional Shr\"odinger
equation with a negative-definite potential which  is known to have a
normalizable bound-state solution with negative $\omega^2$. Due to 
continuity
this bound-state solution should persist for a finite range of
non-vanishing $\kappa$'s. Thus, $\chi$ becomes tachionic, marking an 
instability
of the $\chi = 0$ solution in the core of the defect.

This means that inside the wall
the $SU(2)$ gauge symmetry is spontaneously broken down to $U(1)$.
Two out of three gluons acquire very large masses of order of $v$.
The third gluon becomes a photon. Two degrees of freedom in the
$\chi^a$ field are eaten up by the Higgs mechanism,
the remaining degree of freedom is neutral. The  ``quarks" 
$\psi_{\alpha}$
have charges $\pm 1/2$ with respect to the surviving photon.

Let us have a closer look at the theory emerging in this way.
 Outside the wall the theory has a wider gauge
invariance, $SU(2)$, and is in the non-Abelian confining phase.
The   $U(1)$ gauge invariance is maintained everywhere -- inside 
and outside the wall. The light degrees of freedom inside the wall
are massless ``quarks", interacting through the 
photon exchange. 
We disregard the non-interacting degrees of freedom.
The theory inside the wall is in the Abelian Coulomb phase.
The photon and the light ``quarks" can not escape in the outside space
because there they become a part of the $SU(2)$ theory with no states lighter 
than $\Lambda$. 
The three-dimensional observer confined inside the wall needs
energies of order $\Lambda$
to be able to feel that his/her Universe is actually embedded
in the four-dimensional world.

Let us parenthetically note that the Abelian Coulomb phase
in 2+1 dimensions confines 
electric charges since the potential grows logarithmically with distance.
The three-dimensional electromagnetic coupling constant
$\alpha$ will be of order of $m$, this is also a typical mass
of the neutral bound states whose size will be of order
$L\sim (\mu m)^{-1/2}$. (Recent work \cite{Vol} discusses
a related issue -- the behavior of the fermion zero modes
trapped in the $(2+1)$-dimensional wall in a delocalized
electromagnetic field dispersed in four dimensions.)

Needless to say that a similar mechanism will work for trapping, say,
$SU(2)$ gluons inside the wall submerged into, say, $SU(3)$ environment,
or in any other problem of this type.

\vspace{0.2cm}

{\em 3. Domain wall in supersymmetric gluodynamics} \hspace{1cm}
Consider the simplest SUSY gauge theory, supersymmetric $SU(2)$ 
gluodynamics.
The Lagrangian of the theory
is 
\beq
{\cal L} = \frac{1}{2g^2}\mbox{Tr}\,
\int d^2\theta W^2 = \frac{1}{g^2} \left\{ 
-\frac{1}{4}G_{\mu\nu}^aG_{\mu\nu}^a
+i{\lambda^{a\dagger}}_{\dot\alpha} \partial_\mu 
({\bar\sigma}^\mu)^{\dot\alpha\alpha} {\lambda_\alpha^a}\right\}\, ,
\label{susyL}
\eeq
where $\lambda$ is the gluino field (in the Weyl representation).  Witten's 
index
of this theory is 2, which means that it has two degenerate supersymmetric 
vacua \cite{Witten1}. The vacua are marked by the order parameter
$\lambda\lambda$ (note that the order parameter is $\lambda^2$ or 
$\lambda^{\dagger 2}$,
not $\lambda\lambda^{\dagger}$). One can always adjust the vacuum angle 
$\theta$ in 
such a way that the corresponding VEV is
\beq
{\rm Tr}\langle \lambda^2\rangle = \Lambda^3 \,\,\, \mbox{or}\,\,\, - \Lambda^3\, ,
\label{op}
\eeq
where $\Lambda$ is the scale parameter of SUSY gluodynamics. The $\theta$
dependence of the condensates can be found on general grounds 
\cite{SV1,SV2},
$$
\langle \lambda^2\rangle = \Lambda^3{\rm e}^{i\theta /2}\, ,
$$
so that at $\theta = 2\pi $ the vacua interchange. In this Letter
we will limit ourselves to $\theta = 0$.

The $Z_4$ symmetry of the model is spontaneously broken by the
gluino condensate (\ref{op}) down to $Z_2$. If this is a conventional 
discrete symmetry breaking, there must exist
a wall interpolating between the two vacua. The explicit construction of the 
wall,
 routinely done in the weak coupling theories, is impossible, however,
since $\lambda^2$ is a composite operator, and we are in the strong coupling 
regime. A way out can be indicated. 

The key point is as follows. Superalgebra of the   model under 
consideration
has a very peculiar central extension which reveals itself only in
the wall-like situations. How $N=1$ SUSY can have a central extension is
explained in detail in Ref. \cite{DS}. Briefly,
there exists a trivially conserved ``current"
$J^{\mu\nu\alpha} =\epsilon^{\mu\nu\alpha\beta}\partial_\beta \lambda^2$; 
the corresponding charge
is an anti-symmetric tensor assuming non-vanishing values
only in the presence of walls. 
More specifically,
\beq
\{ Q^\dagger_{\dot\alpha}Q^\dagger_{\dot\beta}\}= \frac{N_c}{4\pi^2}
\left(\Sigma^i\right)_{\dot\alpha\dot\beta}\int \, d^3 x \, \partial_i
({\rm Tr}\lambda\lambda ) \, ;
\label{cext}
\eeq
here $Q^\dagger$ is the supercharge, and $\Sigma^i =\sigma^i\sigma^2$
where $\sigma^i$ stands for the Pauli matrices. Equation
(\ref{cext}) is a quantum anomaly which will be
discussed in more detail in subsequent publications.
For completeness we give here the result referring to
 the gauge group
$SU(N_c)$, rather than to $SU(2)$.

In the true vacuum, when all excitations are localized,
the integral on the right-hand side vanishes identically.
On the wall it reduces
to
$$
qV_2 = \frac{1}{2\pi^2} V_2{\rm Tr} ( \lambda \lambda )\, ,
$$
(for $SU(2)$),
i.e. a non-vanishing central charge emerges.
In the sector with the given (non-zero) value of the central charge
the masses of all states can be shown to satisfy a  quantum Bogomol'nyi
bound,
\beq
M \geq qV_2\, .
\label{BB}
\eeq
The lower bound is achieved on the BPS-saturated states \cite{WO,BPS}.
The domain wall in supersymmetric gluodynamics
has to be such a state. Although we have no rigorous proof
of this statement, we see no reason which could prevent the
BPS saturation. Moreover, in the weakly coupled models
with the non-vanishing central charge one can easily
verify that the wall is a BPS-saturated state.
See e.g. the so-called minimal wall in Ref. \cite{DS}.
For the minimal wall 
the equlity between $\varepsilon$ and the central charge is explicitely 
derived in
\cite{DS}. (Similar relation was first observed in a two-dimensional 
model in \cite{WO}). For the BPS-saturated states one half of the
supersymmetry transformations act trivially, i.e. supersymmetry is
preserved.

The only scenario with no BPS saturated states is when
supersymmetry is completely broken by the solution. Although we mention
this possibility, it is hard to imagine it is realized in SUSY gluodynamics.
Additional arguments in favor of the BPS-saturated wall
are provided by consideration of the weakly coupled SQCD
plus holomorphy arguments, see below.

If the wall in SUSY gluodynamics is the BPS-saturated state
(with half of SUSY transformations acting trivially), the
energy density of the wall is nothing but the value of the central charge 
$q$,
\beq
\varepsilon = |q| = \frac{1}{2\pi^2}|{\rm Tr} \langle \lambda \lambda\rangle |\, .
\label{cech}
\eeq
It remains to be added that the gluino condensate
in supersymmetric gluodynamics 
was exactly calculated in Ref. \cite{SV1} for unitary
and orthogonal groups, and in Ref. \cite{MOS} for all other groups.
Moreover, the wall profile of the order parameter $\lambda\lambda$
is related to the Lagrangian density,
$$
\partial_z (\lambda\lambda)
=G_{\alpha\beta}G_{\alpha\beta} +i\lambda^\dagger_{\dot\alpha}
\partial^{\dot\alpha\beta}\lambda_\beta\, .
$$

For arbitrary gauge group there are $(1/2) T(G)[T(G) -1]$ different walls,
whose energy densities reduce to
$$
\varepsilon = \frac{N_c}{8\pi^2}
|{\rm e}^{2\pi i k/T(G)} - {\rm e}^{2\pi i \ell/T(G)} |
|\langle \lambda \lambda\rangle |\, ,
$$
where $k$ and $\ell$ are integers running from $0$ to $T(G) - 1$. 
The fact that $\varepsilon$ is proportional to the
gluino condensate is not surprising by itself, since both quantities
are of order of $\Lambda^3$. What is non-trivial is the exact
proportionality coefficient. Inside the wall there lives a massless
composite boson and a massless composite fermion; a half of 
supersymmetry is preserved.

If one does not want to rely on the previous results on the gluino condensate
one can calculate directly $\varepsilon$ 
using the same
strategy as was first suggested in Ref. \cite{SV1}
for exact calculation of the gluino condensate. Namely,
one additional quark flavor is  introduced, with the mass term
$m$.  Thus, the original supersymmetric gluodynamics is
substituted by SQCD with one flavor. If the scale parameter
of  SQCD with one flavor is $\tilde\Lambda$ and $m\ll \tilde\Lambda$,
the theory turns out to be in the weakly coupled Higgs phase 
\cite{ADS1}. Then the construction of the domain wall can be carried out,
and $\varepsilon$ calculated. The result depends on the bare mass parameter
$m$ in a holomorphic way; therefore, the exact $m$ dependence can be found,
much in the same way as in Refs.
\cite{SV1,SV2}. Then we can tend $m\ra\infty$, making the matter fields
very heavy. If they are very heavy, they can be integrated out,
and we return back to SUSY gluodynamics. The result for $\varepsilon$
will  still be valid. The last step to be done is to 
express the parameters of SQCD with one light flavor in terms
 of  parameters relevant to SUSY gluodynamics.

Let us outline some basic elements of the procedure.
The structure of the $SU(2)$ model with one flavor is exhaustively described
in the review \cite{VZS}; the reader is referred to this paper
for details and definitions.
One flavor is comprised of two chiral superfields,
$S^{\alpha f}$, where $\alpha$ is the $SU(2)$ index, while $f$ is the subflavor 
index, $f=1,2$. 
 Classically the model has a one-dimensional
vacuum valley (flat direction) parametrized by VEV of the
composite operator $S^2 =S^{\alpha f}S_{\alpha f}$. Quantum-mechanically,
 a superpotential is generated along the valley \cite{ADS1}; it has
 the form
\beq
{\cal W}= \frac{\tilde\Lambda^5}{S^2}
+ \frac{m}{4} S^2\, ,
\label{instspot}
\eeq
where we have included also the (tree level) mass term. The latter stabilizes 
the theory eliminating the run-away vacuum. Note that the mass term {\it
explicitly} breaks the original continuous $R$ symmetry of the theory
down to $Z_2$, under which $S^2 \ra - S^2$. It is a spontaneous breakdown
of this discrete subgroup in the vacuum that gives rise to a domain wall 
solution.
If $m$ is small,
the fields residing in $S^2$ are light, the vacuum expectation value of $S^2$
is large, the $SU(2)$ symmetry (as well as $Z_2$)
is spontaneously broken, the gluons
acquire a large mass (so that they are actually $W$ bosons) and can be 
integrated over. As a  result of this integration, the superpotential
in Eq. (\ref{instspot}) is generated through instantons. Equation 
(\ref{instspot})
is exact -- it has neither perturbative nor non-perturbative corrections
\cite{ADS1,SV1}. The vacuum expectation values of $S^2$ are
\beq
\langle S^2\rangle =  2\tilde\Lambda^{5/2}m^{-1/2} \,\,\, \mbox{or}\,\,\,
- 2\tilde\Lambda^{5/2}m^{-1/2}\, .
\label{VEV}
\eeq

The low-energy theory is that of one chiral superfield; it resembles
the Wess-Zumino model \cite{WZ}. The only difference is a slightly unusual
form of the superpotential, but its particular form  is
unimportant for our purposes. The domain wall in the Wess-Zumino model 
was discussed in 
detail in Ref. \cite{DS} (the ``minimal wall"). In the case at hand it is possible, 
in principle to find an explicit solution interpolating between
two vacua of Eq. (\ref{VEV}), as a function of $z$, which will be valid
almost everywhere. A small interval,  of the size of order $\tilde\Lambda^{-
1}$, 
near the origin where the value of $S^2$ is small, is a strong coupling region.
The semiclassical description of the wall in terms of one superfield $S^2$ is 
invalid here,
since in this region  excitations corresponding to  composite gauge invariant 
operators generically have masses of order $\tilde\Lambda$. The correct 
description
requires many degrees of freedom. Outside the above  narrow region  the wall 
is properly described 
semiclassically by a profile of $S^2$. The width of the wall in
the $z$ direction 
is of order $m^{-1}$. Thus, our wall is a two-component 
construction.
 Our ignorance of the small central region 
does not preclude us from calculating the wall energy density $\varepsilon$
exactly, provided that the description is
continuous. Indeed, the central charge $q$ is related to the
integral
$$
J^{\mu\nu} =\int d^3 x \varepsilon^{0\mu\nu\rho} \partial_\rho  {\cal 
W}(S^2)
\ra \left[ {\cal W}(S^2)_{z\ra +\infty} - {\cal W}(S^2)_{z\ra -\infty}
\right] \, .
$$
At large separations from the wall we are approaching the
true vacuum, where the theory is in the weakly coupled Higgs phase,
and the values of ${\cal W}(S^2)_{z\ra \pm\infty}$ are exactly known.
In this way we find that
$$
q = \mbox{a known const.}\,\times \tilde\Lambda^{5/2}m^{1/2}\, .
$$
Moreover, the condition that 1/2 of SUSY transformations
act trivially (equivalent to the BPS saturation) is explicitely satisfied 
everywhere
in the weak coupling region. By holomorphy we conclude it
must be satisfied in the central region as well. 

Tending $m\ra\infty$ and eliminating $\tilde\Lambda$
in favor of $\Lambda$, the scale parameter of supersymmetric gluodynamics,
we reproduce Eq. (\ref{cech}).

\vspace{0.2cm}
{\em Conclusions} \hspace{1cm} The peculiar features of
the domain wall discussed above are due to
intricacies of the non-Abelian gauge dynamics.
The confining
property of gluodynamics and the dynamical mass gap generation is
the basic element of the
mechanism we suggested  for trapping   massless gauge bosons
inside  the wall. A task that lies ahead is  using this mechanism  in the context 
of
the dynamic compactification scenarios outlined in \cite{DS}. 

As it happened more than once in the past, miracles of supersymmetric 
($N=1$) gauge dynamics allowed us to exactly calculate 
 the energy density of the supersymmetric wall   in the strong coupling 
regime. This new  example of non-trivial physical quantity
that can be found exactly by exploiting specific properties of supersymmetry
is interesting by itself. Moreover, it  settles the issue
of  spontaneous breaking of $Z_{2T(G)}$ in supersymmetric gluodynamics
{\em versus} a new superselection rule, 
which was debated over a decade, in favor of the first option.

\vspace{0.3cm}

{\em Acknowledgments} \hspace{1cm} The authors
would like to thank I. Kogan,  C. Korthals-Altes,   T. Ortin,  F. Quevedo, and  A. 
Smilga
for useful comments.

This work was supported in part by DOE under the grant number
DE-FG02-94ER40823.

\end{document}